\documentstyle[12pt]{article}
\newcommand{\be}{\begin{eqnarray}}
\newcommand{\ee}{\end{eqnarray}}
\newcommand{\beq}{\begin{equation}}
\newcommand{\eeq}{\end{equation}}
\begin{document}
%%%%%%%%%%%%%%%%%%%%%%%%%%%%%%%%%%%%%%%%%%%%%%%%%%%%%%%%%%%%%%%%%%%%%%%%%%%%%%
\begin{flushright}
DFTT 44/96 \\
IFT-96-16\\
MPI-PhT/96-63\\
hep-ph/9607427 \\
\end{flushright}
\vskip 1.5cm
\begin{center}
{\bf New proton polarized structure functions\\ in 
charged current processes at HERA}\\
\vskip 1.5cm
{\sf M.\ Anselmino$^*$, P. Gambino$^{**}$, J. Kalinowski$^{***,}$}\footnote{
Supported in part by the Polish Committee for Scientific Research.}
\vskip 0.8cm
{$^*$Dipartimento di Fisica Teorica, Universit\`a di Torino and \\
      INFN, Sezione di Torino, Via P. Giuria 1, 10125 Torino, Italy\\
\vskip 0.5cm
$^{**}$Max Planck Institut f\"ur Physik, Werner Heisenberg Institut, \\
F\"ohringer Ring 6, D80805 M\"unchen, Germany \\
\vskip 0.5cm
$^{***}$Warsaw University, Institute of Theoretical Physics, \\
Ul. Hoza 69, 00681 Warsaw, Poland }\\
\end{center}
\vskip 1.5cm
\noindent
{\bf Abstract:} \\
Estimates for longitudinal spin asymmetries which single out 
new polarized nucleon structure functions in deeply inelastic charged 
current interactions at HERA energies are given, exploiting their 
interpretation in terms of polarized quark distributions. 
These asymmetries turn out to be large and allow a measurement of the 
new polarized structure functions $g_1^{W}$ and $g_5^{W}$, which 
would add valuable tests and information on the spin content of quarks
inside a polarized proton. We also show that single spin asymmetries
in neutral current interactions are very small.

%%%%%%%%%%%%%%%%%%%%%%%%%%%%%%%%%%%%%%%%%%%%%%%%%%%%%%%%%%%%%%%%%%%%%%%%%%%%%%%
\newpage
%%%%%%%%%%%%%%%%%%%%%%%%%%%%%%%%%%%%%%%%%%%%%%%%%%%%%%%%%%%%%%%%%%%%%%%%%%%%%%%
\pagestyle{plain}
The possible availability of a polarized proton beam at HERA, with 
polarized lepton-proton scattering at $Q^2$ values up to a few thousands of
GeV$^2$, allows a new investigation of proton spin effects in 
charged current ($cc$) interactions \cite{hera}. This gives access to 
new polarized proton structure functions which would offer novel information 
on polarized quark distributions, because charged current events probe 
combinations of spin-dependent quark distributions different from those probed 
in electromagnetic processes \cite{ael}-\cite{bk}. It is then clear the 
importance of such future HERA spin program.

We give here simple estimates of double longitudinal spin asymmetries in 
$\ell^{\mp} p \to \nu (\bar\nu) X$ processes which, if measured, would yield 
direct information on new structure functions, including polarized parity 
violating ones. It turns out that such asymmetries are large, much larger than 
the expected statistical errors, and can be safely measured, provided charged
current events may be detected, which is expected. More spin asymmetries and 
their content in terms of structure functions and quark distributions are 
discussed in Ref. \cite{agk}.

According to the conventions and notations of Ref. \cite{agk}, the 
cross-section for the weak interaction of a longitudinally polarized lepton 
(with helicity $\lambda/2$) with a longitudinally polarized nucleon (with spin 
four-vector $S=S_L$ ) in the charged current process $\ell^{\mp} p \to 
\nu (\bar\nu) X$ is given by (neglecting terms of order $m_N/E_N$):
\be
{d^2\sigma_{cc}^{\ell^{\mp}p} \over dx \, dQ^2} (\lambda, S_L) &=&
{G^2 M_W^4 \over 4\pi} {(1 \mp \lambda) \over (Q^2 + M_W^2)^2}
\biggl\{ (y^2-2y+2) F_1^{W^{\mp}}
- \lambda \, {y(2-y) \over 2} F_3^{W^{\mp}} \nonumber \\
&-& \lambda y(2-y) \, g_1^{W^{\mp}} + (y^2-2y+2) \, g_5^{W^{\mp}} \biggr\} \,.
\label{sig} 
\ee
Notice that this implies $\lambda = -1$ for negatively charged leptons,
$\ell^-$, and $\lambda = +1$  for positively charged ones, $\ell^+$.
When reversing the nucleon spin ($S_L \to -S_L$) the terms in $g_1^W$
and $g_5^W$ change sign. 

By setting
\beq
a \equiv 2(y^2-2y+2) \quad\quad\quad b \equiv y(2-y) \,,
\eeq
Eq. (\ref{sig}) can be written in a simple form as
\beq
{d^2\sigma_{cc}^{\ell^{\mp} p} \over dx \, dQ^2} (\lambda = \mp1, S_L)
= {1\over 4\pi} {G^2 M_W^4 \over (Q^2 + M_W^2)^2}
\left\{ a \left[F_1^{W^{\mp}} + g_5^{W^{\mp}}\right]
\pm b \left[F_3^{W^{\mp}} + 2 g_1^{W^{\mp}}\right] \right\},
\label{sigs}
\eeq
where the $\mp$ refer respectively to $\ell^-$ and $\ell^+$. In case the 
leptons were not 100\% polarized (that is $\langle \lambda \rangle \not= 
\mp 1$) the r.h.s. of the above equation should be multiplied by an overall
factor $(1 \mp \langle \lambda \rangle)/2$; 
for unpolarized leptons ($\langle \lambda \rangle = 0$) this   
amounts to a factor 1/2.

It is now immediate to compute the double spin asymmetry
\beq
A^{W^{\mp}} \equiv {d^2 \sigma_{cc}^{\ell^{\mp} p} (\mp, S_L) - 
d^2 \sigma_{cc}^{\ell^{\mp} p} (\mp, - S_L)
\over d^2 \sigma_{cc}^{\ell^{\mp} p} (\mp, S_L) + 
d^2 \sigma_{cc}^{\ell^{\mp} p} (\mp, - S_L)} 
= {\pm 2 b \, g_1^{W^{\mp}} + a \, g_5^{W^{\mp}} \over 
a \, F_1^{W^{\mp}}  \pm b \, F_3^{W^{\mp}}} \, \cdot
\label{asy} 
\eeq
Notice that the same result holds with unpolarized leptons, as the $cc$
interactions single out definite lepton helicities.

Recalling the quark parton model expression of the structure functions
\cite{agk},
\be
g_1^{W^-} = \Delta u + \Delta c + \Delta\bar d + \Delta\bar s \quad\quad 
F_1^{W^-} &=& u + c + \bar d + \bar s \\
g_5^{W^-} = \Delta u + \Delta c - \Delta\bar d - \Delta\bar s \quad\quad 
F_3^{W^-} &=& 2(u + c - \bar d - \bar s) \\
g_1^{W^+} = \Delta d + \Delta s + \Delta\bar u + \Delta\bar c \quad\quad 
F_1^{W^+} &=& d + s + \bar u + \bar c \\
g_5^{W^+} = \Delta d + \Delta s - \Delta\bar u - \Delta\bar c \quad\quad 
F_3^{W^+} &=& 2(d + s - \bar u - \bar c) \,,
\ee
one has, respectively for charged current processes initiated by 
negatively or positively charged leptons:
\beq
A^{W^-} = {(\Delta u + \Delta c) - (1-y)^2 (\Delta\bar d + \Delta\bar s) 
\over (u + c) + (1-y)^2 (\bar d + \bar s)} 
\eeq
and
\beq
A^{W^+} = {(1-y)^2 (\Delta d + \Delta s) - (\Delta\bar u + \Delta\bar c) 
\over (1-y)^2 (d + s) + (\bar u + \bar c)} \, \cdot
\eeq

Other interesting combinations of polarized quark distribution functions 
may be obtained if one could combine data taken with $\ell^-$ and 
$\ell^+$ beams. This is discussed in detail in Ref. \cite{agk}. 

Notice that, in principle, one might easily extract
information on parity violating polarized structure functions also by 
considering a single spin asymmetry in neutral current ($nc$) processes 
with longitudinally polarized protons and unpolarized leptons \cite{agk,bil}.
The spin asymmetry for unpolarized electrons reads \cite{agk}
\beq
A^{e^- p}_{nc}=
\eta^{\gamma Z} \>
\frac{a \left[ (2 s_w^2-\frac1{2}) g_5^{\gamma Z} + 
\eta^{\gamma Z}(4 s^4_w - 2 s^2_w + \frac1{2}) g_5^Z) \right]
-  b \left[ g_1^{\gamma Z} + \eta^{\gamma Z} (4 s^2_w -1) g_1^Z \right] }{
a \left[ F_1^{\gamma } + \eta^{\gamma Z} (2 s_w^2-\frac1{2}) F_1^{\gamma Z} 
\right] - \frac1{2} b \, \eta^{\gamma Z} \, F_3^{\gamma Z} 
+ {\cal O}[(\eta^{\gamma Z})^2]} \,,
\eeq
 where   
\beq \eta^{\gamma Z}=\frac{GM_Z^2}{2\sqrt{2}\pi\alpha}
\frac{Q^2}{Q^2+M_Z^2}\> ,
\eeq
$s_w$ is the  sine of the Weinberg angle, and we have neglected in the 
denominator the purely weak contributions, of ${\cal O}[(\eta^{\gamma Z})^2]$ 
and negligible in most of HERA $Q^2$ range where $\eta^{\gamma Z}$ is still 
rather small. The expression of the structure functions $g^{\gamma Z}$ and 
$g^Z$ in terms of quark distributions can be found in Ref. \cite{agk}. 

In  the numerical analysis we use a lepton energy $E_\ell=27.6$ GeV 
and a proton beam energy of 820 GeV and impose the following kinematical cuts:
$0.01 < y < 0.9$, $x<0.7$. In addition, in the $cc$ case we
apply a cut on the missing transverse momentum, $\not\hskip -1.5mm P_T>15$ GeV,
and in the $nc$ case we demand $Q^2 > 500$ GeV$^2$. The leading order 
polarized parton distributions are taken from Ref. \cite{polpar} 
(here we use the set ``standard scenario'', eventually extrapolated to $Q^2$ 
values greater than $10^4$ GeV$^2$; the  ``valence" set yields very similar  
results); the unpolarized ones are taken from Ref. \cite{unpar}. 
We have assumed fully polarized beams, but we emphasize that $A^{W^\mp}$ 
are independent of the degree of polarization of the lepton beam, 
and proportional to the polarization of the proton: 
$|\langle\lambda_\ell \rangle|<1$ would only marginally decrease the available
statistical precision, while $|\langle\lambda_p \rangle|<1$ would simply
reduce the asymmetries by the  corresponding factor. A high degree 
of polarization of the proton beam, more than that of the leptons,
would be a most welcome feature.

In Figs. 1 -- 6 we show the expected asymmetries as functions of $x$ both 
at fixed $Q^2=1000$ GeV$^2$ and averaging over all the kinematically 
allowed $Q^2$ region, that is summing all detectable events in a given
$x$-bin.  
The statistical errors shown in the figures have been calculated
from 
\beq \delta A=\frac{\sqrt{1-A^2}}{\sqrt{2{\cal L}(d^2\sigma/dx dQ^2) \Delta x
\Delta Q^2}}
\eeq
assuming that for each longitudinal proton polarization an integrated 
luminosity  ${\cal L}=500$ pb$^{-1}$ can be collected. [To calculate 
the errors at fixed $Q^2$ we take $\Delta Q^2=Q^2/2$.] It is interesting 
to note that for relatively low $Q^2$ ($\approx 10^3$ GeV$^2$)
and in the kinematically allowed region, $a\gg b$, so that the asymmetries 
are dominated by the parity violating form factor $g_5$. This hints to 
the possibility of a direct extraction of this interesting quantity from 
the asymmetry.

We find that in the $cc$ case the spin asymmetries, both in $e^-p$ and $e^+p$ 
processes, are large (up to 60\%) and measurable in an upgraded
HERA experiment. As the statistical errors scale like $1/\sqrt{{\cal L}}$, 
a reasonable precision can be achieved even with an integrated luminosity  
of only ${\cal L}=50$ pb$^{-1}$ per polarization. On the other hand, 
in the $nc$ case the spin asymmetry is much too small to be 
meaningfully measured at HERA, as an effect of the 
dominant parity conserving electromagnetic contribution.

In summary, 
we have clearly shown that measurements of charged current processes
in deep inelastic scattering of longitudinally polarized or unpolarized 
leptons off longitudinally polarized protons at HERA energies offer a 
unique and viable way of reaching more information on the quark spin 
content of protons; predictions based on the actual knowledge of the 
polarized quark distributions are given. 

\vskip .8cm
Upon completion of this analysis we learned that similar conclusions for 
the spin asymmetries have been obtained in Ref. \cite{maul}
%\vskip 1cm

%\end{document}
\newpage 
\setlength\textheight{23.0cm}
\setlength\textwidth{16.5cm}
\setlength\oddsidemargin{-0.1cm}
\setlength\evensidemargin{-0.1cm}
\headsep 30pt
\begin{figure}
% GNUPLOT: LaTeX picture
\setlength{\unitlength}{0.240900pt}
\ifx\plotpoint\undefined\newsavebox{\plotpoint}\fi
\sbox{\plotpoint}{\rule[-0.175pt]{0.350pt}{0.350pt}}%
\begin{picture}(1559,1000)(0,0)
\sbox{\plotpoint}{\rule[-0.175pt]{0.350pt}{0.350pt}}%
%\put(264,158){\rule[-0.175pt]{296..548pt}{0.350pt}}
\put(264,158){\rule[-0.175pt]{4.818pt}{0.350pt}}
\put(242,158){\makebox(0,0)[r]{0}}
\put(1475,158){\rule[-0.175pt]{4.818pt}{0.350pt}}
\put(264,255){\rule[-0.175pt]{4.818pt}{0.350pt}}
\put(242,255){\makebox(0,0)[r]{}}
\put(1475,255){\rule[-0.175pt]{4.818pt}{0.350pt}}
\put(264,353){\rule[-0.175pt]{4.818pt}{0.350pt}}
\put(242,353){\makebox(0,0)[r]{0.2}}
\put(1475,353){\rule[-0.175pt]{4.818pt}{0.350pt}}
\put(264,450){\rule[-0.175pt]{4.818pt}{0.350pt}}
\put(242,450){\makebox(0,0)[r]{}}
\put(1475,450){\rule[-0.175pt]{4.818pt}{0.350pt}}
\put(264,547){\rule[-0.175pt]{4.818pt}{0.350pt}}
\put(242,547){\makebox(0,0)[r]{0.4}}
\put(1475,547){\rule[-0.175pt]{4.818pt}{0.350pt}}
\put(264,644){\rule[-0.175pt]{4.818pt}{0.350pt}}
\put(242,644){\makebox(0,0)[r]{}}
\put(1475,644){\rule[-0.175pt]{4.818pt}{0.350pt}}
\put(264,742){\rule[-0.175pt]{4.818pt}{0.350pt}}
\put(242,742){\makebox(0,0)[r]{0.6}}
\put(1475,742){\rule[-0.175pt]{4.818pt}{0.350pt}}
\put(264,839){\rule[-0.175pt]{4.818pt}{0.350pt}}
\put(242,839){\makebox(0,0)[r]{}}
\put(1475,839){\rule[-0.175pt]{4.818pt}{0.350pt}}
\put(264,936){\rule[-0.175pt]{4.818pt}{0.350pt}}
\put(242,936){\makebox(0,0)[r]{0.8}}
\put(1475,936){\rule[-0.175pt]{4.818pt}{0.350pt}}
\put(264,158){\rule[-0.175pt]{0.350pt}{4.818pt}}
\put(264,113){\makebox(0,0){0.01}}
\put(264,916){\rule[-0.175pt]{0.350pt}{4.818pt}}
\put(449,158){\rule[-0.175pt]{0.350pt}{2.409pt}}
\put(449,926){\rule[-0.175pt]{0.350pt}{2.409pt}}
\put(558,158){\rule[-0.175pt]{0.350pt}{2.409pt}}
\put(558,926){\rule[-0.175pt]{0.350pt}{2.409pt}}
\put(635,158){\rule[-0.175pt]{0.350pt}{2.409pt}}
\put(635,926){\rule[-0.175pt]{0.350pt}{2.409pt}}
\put(694,158){\rule[-0.175pt]{0.350pt}{2.409pt}}
\put(694,926){\rule[-0.175pt]{0.350pt}{2.409pt}}
\put(743,158){\rule[-0.175pt]{0.350pt}{2.409pt}}
\put(743,926){\rule[-0.175pt]{0.350pt}{2.409pt}}
\put(784,158){\rule[-0.175pt]{0.350pt}{2.409pt}}
\put(784,926){\rule[-0.175pt]{0.350pt}{2.409pt}}
\put(820,158){\rule[-0.175pt]{0.350pt}{2.409pt}}
\put(820,926){\rule[-0.175pt]{0.350pt}{2.409pt}}
\put(851,158){\rule[-0.175pt]{0.350pt}{2.409pt}}
\put(851,926){\rule[-0.175pt]{0.350pt}{2.409pt}}
\put(880,158){\rule[-0.175pt]{0.350pt}{4.818pt}}
\put(880,113){\makebox(0,0){0.1}}
\put(880,916){\rule[-0.175pt]{0.350pt}{4.818pt}}
\put(1065,158){\rule[-0.175pt]{0.350pt}{2.409pt}}
\put(1065,926){\rule[-0.175pt]{0.350pt}{2.409pt}}
\put(1173,158){\rule[-0.175pt]{0.350pt}{2.409pt}}
\put(1173,926){\rule[-0.175pt]{0.350pt}{2.409pt}}
\put(1250,158){\rule[-0.175pt]{0.350pt}{2.409pt}}
\put(1250,926){\rule[-0.175pt]{0.350pt}{2.409pt}}
\put(1310,158){\rule[-0.175pt]{0.350pt}{2.409pt}}
\put(1310,926){\rule[-0.175pt]{0.350pt}{2.409pt}}
\put(1358,158){\rule[-0.175pt]{0.350pt}{2.409pt}}
\put(1358,926){\rule[-0.175pt]{0.350pt}{2.409pt}}
\put(1400,158){\rule[-0.175pt]{0.350pt}{2.409pt}}
\put(1400,926){\rule[-0.175pt]{0.350pt}{2.409pt}}
\put(1435,158){\rule[-0.175pt]{0.350pt}{2.409pt}}
\put(1435,926){\rule[-0.175pt]{0.350pt}{2.409pt}}
\put(1467,158){\rule[-0.175pt]{0.350pt}{2.409pt}}
\put(1467,926){\rule[-0.175pt]{0.350pt}{2.409pt}}
\put(1495,158){\rule[-0.175pt]{0.350pt}{4.818pt}}
\put(1495,113){\makebox(0,0){1}}
\put(1495,916){\rule[-0.175pt]{0.350pt}{4.818pt}}
\put(264,158){\rule[-0.175pt]{296.548pt}{0.350pt}}
\put(1495,158){\rule[-0.175pt]{0.350pt}{187.420pt}}
\put(264,936){\rule[-0.175pt]{296.548pt}{0.350pt}}
\put(45,547){\makebox(0,0)[l]{\shortstack{$ A^{W^-}$}}}
\put(879,68){\makebox(0,0){$x$}}
\put(264,158){\rule[-0.175pt]{0.350pt}{187.420pt}}
\put(1365,871){\makebox(0,0)[r]{$A^{W^-}$}}
\put(1409,871){\raisebox{-1.2pt}{\makebox(0,0){$\Diamond$}}}
\put(387,311){\raisebox{-1.2pt}{\makebox(0,0){$\Diamond$}}}
\put(510,337){\raisebox{-1.2pt}{\makebox(0,0){$\Diamond$}}}
\put(633,374){\raisebox{-1.2pt}{\makebox(0,0){$\Diamond$}}}
\put(756,427){\raisebox{-1.2pt}{\makebox(0,0){$\Diamond$}}}
\put(880,493){\raisebox{-1.2pt}{\makebox(0,0){$\Diamond$}}}
\put(1003,568){\raisebox{-1.2pt}{\makebox(0,0){$\Diamond$}}}
\put(1126,639){\raisebox{-1.2pt}{\makebox(0,0){$\Diamond$}}}
\put(1249,696){\raisebox{-1.2pt}{\makebox(0,0){$\Diamond$}}}
\put(1372,741){\raisebox{-1.2pt}{\makebox(0,0){$\Diamond$}}}
\put(1387,871){\rule[-0.175pt]{15.899pt}{0.350pt}}
\put(1387,861){\rule[-0.175pt]{0.350pt}{4.818pt}}
\put(1453,861){\rule[-0.175pt]{0.350pt}{4.818pt}}
\put(387,290){\rule[-0.175pt]{0.350pt}{10.118pt}}
\put(377,290){\rule[-0.175pt]{4.818pt}{0.350pt}}
\put(377,332){\rule[-0.175pt]{4.818pt}{0.350pt}}
\put(510,317){\rule[-0.175pt]{0.350pt}{9.636pt}}
\put(500,317){\rule[-0.175pt]{4.818pt}{0.350pt}}
\put(500,357){\rule[-0.175pt]{4.818pt}{0.350pt}}
\put(633,355){\rule[-0.175pt]{0.350pt}{9.395pt}}
\put(623,355){\rule[-0.175pt]{4.818pt}{0.350pt}}
\put(623,394){\rule[-0.175pt]{4.818pt}{0.350pt}}
\put(756,408){\rule[-0.175pt]{0.350pt}{9.154pt}}
\put(746,408){\rule[-0.175pt]{4.818pt}{0.350pt}}
\put(746,446){\rule[-0.175pt]{4.818pt}{0.350pt}}
\put(880,473){\rule[-0.175pt]{0.350pt}{9.395pt}}
\put(870,473){\rule[-0.175pt]{4.818pt}{0.350pt}}
\put(870,512){\rule[-0.175pt]{4.818pt}{0.350pt}}
\put(1003,548){\rule[-0.175pt]{0.350pt}{9.636pt}}
\put(993,548){\rule[-0.175pt]{4.818pt}{0.350pt}}
\put(993,588){\rule[-0.175pt]{4.818pt}{0.350pt}}
\put(1126,618){\rule[-0.175pt]{0.350pt}{10.359pt}}
\put(1116,618){\rule[-0.175pt]{4.818pt}{0.350pt}}
\put(1116,661){\rule[-0.175pt]{4.818pt}{0.350pt}}
\put(1249,668){\rule[-0.175pt]{0.350pt}{13.490pt}}
\put(1239,668){\rule[-0.175pt]{4.818pt}{0.350pt}}
\put(1239,724){\rule[-0.175pt]{4.818pt}{0.350pt}}
\put(1372,680){\rule[-0.175pt]{0.350pt}{29.149pt}}
\put(1362,680){\rule[-0.175pt]{4.818pt}{0.350pt}}
\put(1362,801){\rule[-0.175pt]{4.818pt}{0.350pt}}
\end{picture}
\caption{ \small\sf
Charged current $e^- p$ spin asymmetry, Eq. (9) of text, 
for $Q^2=1000$ GeV$^2$, at HERA energies with fully
polarized beams. Statistical errors calculated for 
${\cal L}= 500 \, pb^{-1}$ and $\Delta Q^2=500$ GeV$^2$
for each polarization are shown.
}
\end{figure}
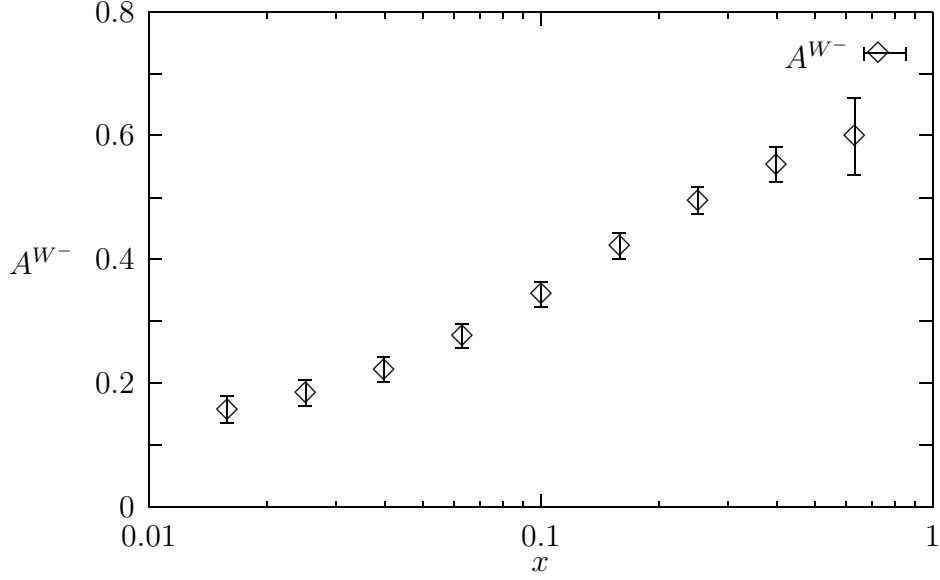
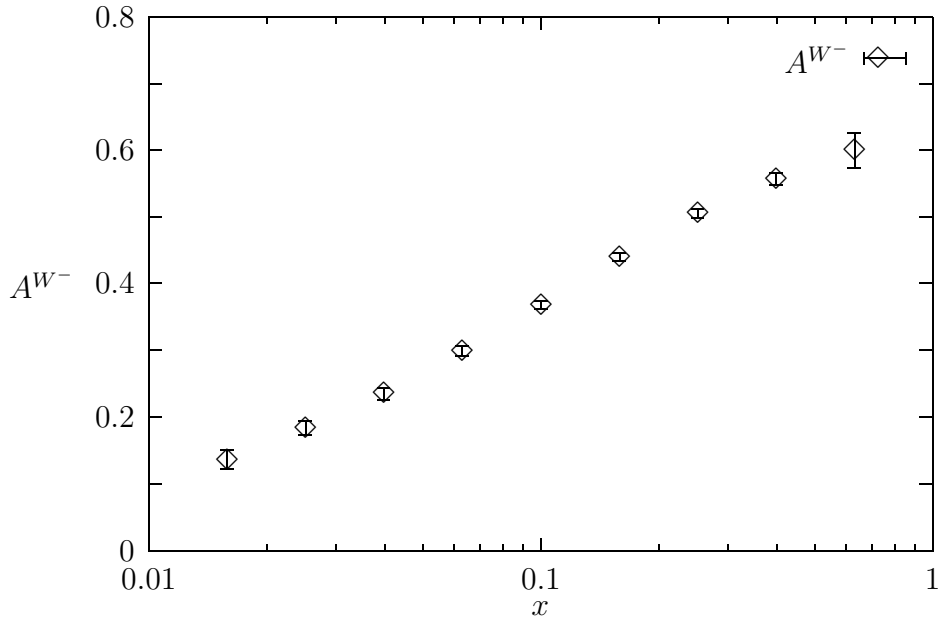
\begin{figure}
% GNUPLOT: LaTeX picture
\setlength{\unitlength}{0.240900pt}
\ifx\plotpoint\undefined\newsavebox{\plotpoint}\fi
\sbox{\plotpoint}{\rule[-0.175pt]{0.350pt}{0.350pt}}%
%\begin{picture}(1559,1110)(0,0)
\begin{picture}(1559,1090)(0,0)
%\tenrm
\sbox{\plotpoint}{\rule[-0.175pt]{0.350pt}{0.350pt}}%
\put(264,158){\rule[-0.175pt]{296.548pt}{0.350pt}}
\put(264,158){\rule[-0.175pt]{4.818pt}{0.350pt}}
\put(242,158){\makebox(0,0)[r]{0}}
\put(1475,158){\rule[-0.175pt]{4.818pt}{0.350pt}}
\put(264,263){\rule[-0.175pt]{4.818pt}{0.350pt}}
\put(242,263){\makebox(0,0)[r]{}}
\put(1475,263){\rule[-0.175pt]{4.818pt}{0.350pt}}
\put(264,368){\rule[-0.175pt]{4.818pt}{0.350pt}}
\put(242,368){\makebox(0,0)[r]{0.2}}
\put(1475,368){\rule[-0.175pt]{4.818pt}{0.350pt}}
\put(264,473){\rule[-0.175pt]{4.818pt}{0.350pt}}
\put(242,473){\makebox(0,0)[r]{}}
\put(1475,473){\rule[-0.175pt]{4.818pt}{0.350pt}}
\put(264,578){\rule[-0.175pt]{4.818pt}{0.350pt}}
\put(242,578){\makebox(0,0)[r]{0.4}}
\put(1475,578){\rule[-0.175pt]{4.818pt}{0.350pt}}
\put(264,682){\rule[-0.175pt]{4.818pt}{0.350pt}}
\put(242,682){\makebox(0,0)[r]{}}
\put(1475,682){\rule[-0.175pt]{4.818pt}{0.350pt}}
\put(264,787){\rule[-0.175pt]{4.818pt}{0.350pt}}
\put(242,787){\makebox(0,0)[r]{0.6}}
\put(1475,787){\rule[-0.175pt]{4.818pt}{0.350pt}}
\put(264,892){\rule[-0.175pt]{4.818pt}{0.350pt}}
\put(242,892){\makebox(0,0)[r]{}}
\put(1475,892){\rule[-0.175pt]{4.818pt}{0.350pt}}
\put(264,997){\rule[-0.175pt]{4.818pt}{0.350pt}}
\put(242,997){\makebox(0,0)[r]{0.8}}
\put(1475,997){\rule[-0.175pt]{4.818pt}{0.350pt}}
\put(264,158){\rule[-0.175pt]{0.350pt}{4.818pt}}
\put(264,113){\makebox(0,0){0.01}}
\put(264,977){\rule[-0.175pt]{0.350pt}{4.818pt}}
\put(449,158){\rule[-0.175pt]{0.350pt}{2.409pt}}
\put(449,987){\rule[-0.175pt]{0.350pt}{2.409pt}}
\put(558,158){\rule[-0.175pt]{0.350pt}{2.409pt}}
\put(558,987){\rule[-0.175pt]{0.350pt}{2.409pt}}
\put(635,158){\rule[-0.175pt]{0.350pt}{2.409pt}}
\put(635,987){\rule[-0.175pt]{0.350pt}{2.409pt}}
\put(694,158){\rule[-0.175pt]{0.350pt}{2.409pt}}
\put(694,987){\rule[-0.175pt]{0.350pt}{2.409pt}}
\put(743,158){\rule[-0.175pt]{0.350pt}{2.409pt}}
\put(743,987){\rule[-0.175pt]{0.350pt}{2.409pt}}
\put(784,158){\rule[-0.175pt]{0.350pt}{2.409pt}}
\put(784,987){\rule[-0.175pt]{0.350pt}{2.409pt}}
\put(820,158){\rule[-0.175pt]{0.350pt}{2.409pt}}
\put(820,987){\rule[-0.175pt]{0.350pt}{2.409pt}}
\put(851,158){\rule[-0.175pt]{0.350pt}{2.409pt}}
\put(851,987){\rule[-0.175pt]{0.350pt}{2.409pt}}
\put(880,158){\rule[-0.175pt]{0.350pt}{4.818pt}}
\put(880,113){\makebox(0,0){0.1}}
\put(880,977){\rule[-0.175pt]{0.350pt}{4.818pt}}
\put(1065,158){\rule[-0.175pt]{0.350pt}{2.409pt}}
\put(1065,987){\rule[-0.175pt]{0.350pt}{2.409pt}}
\put(1173,158){\rule[-0.175pt]{0.350pt}{2.409pt}}
\put(1173,987){\rule[-0.175pt]{0.350pt}{2.409pt}}
\put(1250,158){\rule[-0.175pt]{0.350pt}{2.409pt}}
\put(1250,987){\rule[-0.175pt]{0.350pt}{2.409pt}}
\put(1310,158){\rule[-0.175pt]{0.350pt}{2.409pt}}
\put(1310,987){\rule[-0.175pt]{0.350pt}{2.409pt}}
\put(1358,158){\rule[-0.175pt]{0.350pt}{2.409pt}}
\put(1358,987){\rule[-0.175pt]{0.350pt}{2.409pt}}
\put(1400,158){\rule[-0.175pt]{0.350pt}{2.409pt}}
\put(1400,987){\rule[-0.175pt]{0.350pt}{2.409pt}}
\put(1435,158){\rule[-0.175pt]{0.350pt}{2.409pt}}
\put(1435,987){\rule[-0.175pt]{0.350pt}{2.409pt}}
\put(1467,158){\rule[-0.175pt]{0.350pt}{2.409pt}}
\put(1467,987){\rule[-0.175pt]{0.350pt}{2.409pt}}
\put(1495,158){\rule[-0.175pt]{0.350pt}{4.818pt}}
\put(1495,113){\makebox(0,0){1}}
\put(1495,977){\rule[-0.175pt]{0.350pt}{4.818pt}}
\put(264,158){\rule[-0.175pt]{296.548pt}{0.350pt}}
\put(1495,158){\rule[-0.175pt]{0.350pt}{202.115pt}}
\put(264,997){\rule[-0.175pt]{296.548pt}{0.350pt}}
\put(45,577){\makebox(0,0)[l]{\shortstack{$A^{W^-}$}}}
\put(879,68){\makebox(0,0){$x$}}
\put(264,158){\rule[-0.175pt]{0.350pt}{202.115pt}}
\put(1365,932){\makebox(0,0)[r]{$A^{W^-}$}}
\put(1409,932){\raisebox{-1.2pt}{\makebox(0,0){$\Diamond$}}}
\put(387,301){\raisebox{-1.2pt}{\makebox(0,0){$\Diamond$}}}
\put(510,351){\raisebox{-1.2pt}{\makebox(0,0){$\Diamond$}}}
\put(633,405){\raisebox{-1.2pt}{\makebox(0,0){$\Diamond$}}}
\put(756,472){\raisebox{-1.2pt}{\makebox(0,0){$\Diamond$}}}
\put(880,544){\raisebox{-1.2pt}{\makebox(0,0){$\Diamond$}}}
\put(1003,620){\raisebox{-1.2pt}{\makebox(0,0){$\Diamond$}}}
\put(1126,688){\raisebox{-1.2pt}{\makebox(0,0){$\Diamond$}}}
\put(1249,742){\raisebox{-1.2pt}{\makebox(0,0){$\Diamond$}}}
\put(1372,788){\raisebox{-1.2pt}{\makebox(0,0){$\Diamond$}}}
\put(1387,932){\rule[-0.175pt]{15.899pt}{0.350pt}}
\put(1387,922){\rule[-0.175pt]{0.350pt}{4.818pt}}
\put(1453,922){\rule[-0.175pt]{0.350pt}{4.818pt}}
\put(387,286){\rule[-0.175pt]{0.350pt}{7.227pt}}
\put(377,286){\rule[-0.175pt]{4.818pt}{0.350pt}}
\put(377,316){\rule[-0.175pt]{4.818pt}{0.350pt}}
\put(510,340){\rule[-0.175pt]{0.350pt}{5.300pt}}
\put(500,340){\rule[-0.175pt]{4.818pt}{0.350pt}}
\put(500,362){\rule[-0.175pt]{4.818pt}{0.350pt}}
\put(633,395){\rule[-0.175pt]{0.350pt}{4.577pt}}
\put(623,395){\rule[-0.175pt]{4.818pt}{0.350pt}}
\put(623,414){\rule[-0.175pt]{4.818pt}{0.350pt}}
\put(756,464){\rule[-0.175pt]{0.350pt}{3.854pt}}
\put(746,464){\rule[-0.175pt]{4.818pt}{0.350pt}}
\put(746,480){\rule[-0.175pt]{4.818pt}{0.350pt}}
\put(880,537){\rule[-0.175pt]{0.350pt}{3.373pt}}
\put(870,537){\rule[-0.175pt]{4.818pt}{0.350pt}}
\put(870,551){\rule[-0.175pt]{4.818pt}{0.350pt}}
\put(1003,613){\rule[-0.175pt]{0.350pt}{3.132pt}}
\put(993,613){\rule[-0.175pt]{4.818pt}{0.350pt}}
\put(993,626){\rule[-0.175pt]{4.818pt}{0.350pt}}
\put(1126,681){\rule[-0.175pt]{0.350pt}{3.373pt}}
\put(1116,681){\rule[-0.175pt]{4.818pt}{0.350pt}}
\put(1116,695){\rule[-0.175pt]{4.818pt}{0.350pt}}
\put(1249,733){\rule[-0.175pt]{0.350pt}{4.336pt}}
\put(1239,733){\rule[-0.175pt]{4.818pt}{0.350pt}}
\put(1239,751){\rule[-0.175pt]{4.818pt}{0.350pt}}
\put(1372,760){\rule[-0.175pt]{0.350pt}{13.249pt}}
\put(1362,760){\rule[-0.175pt]{4.818pt}{0.350pt}}
\put(1362,815){\rule[-0.175pt]{4.818pt}{0.350pt}}
\end{picture}
\caption{\small\sf 
Charged current $e^- p$ spin asymmetry, Eq. (9) of text, 
averaged over the allowed $Q^2$ region, at HERA energies with fully
polarized beams. Statistical errors calculated for 
${\cal L}= 500 \, pb^{-1}$ for each polarization are shown.
}
\end{figure}
\begin{figure}
% GNUPLOT: LaTeX picture
\setlength{\unitlength}{0.240900pt}
\ifx\plotpoint\undefined\newsavebox{\plotpoint}\fi
\sbox{\plotpoint}{\rule[-0.175pt]{0.350pt}{0.350pt}}%
\begin{picture}(1559,1049)(0,0)
\sbox{\plotpoint}{\rule[-0.175pt]{0.350pt}{0.350pt}}%
\put(264,825){\rule[-0.175pt]{296.548pt}{0.350pt}}
\put(264,158){\rule[-0.175pt]{4.818pt}{0.350pt}}
\put(242,158){\makebox(0,0)[r]{-0.6}}
\put(1475,158){\rule[-0.175pt]{4.818pt}{0.350pt}}
\put(264,269){\rule[-0.175pt]{4.818pt}{0.350pt}}
\put(242,269){\makebox(0,0)[r]{-0.5}}
\put(1475,269){\rule[-0.175pt]{4.818pt}{0.350pt}}
\put(264,380){\rule[-0.175pt]{4.818pt}{0.350pt}}
\put(242,380){\makebox(0,0)[r]{-0.4}}
\put(1475,380){\rule[-0.175pt]{4.818pt}{0.350pt}}
\put(264,491){\rule[-0.175pt]{4.818pt}{0.350pt}}
\put(242,491){\makebox(0,0)[r]{-0.3}}
\put(1475,491){\rule[-0.175pt]{4.818pt}{0.350pt}}
\put(264,603){\rule[-0.175pt]{4.818pt}{0.350pt}}
\put(242,603){\makebox(0,0)[r]{-0.2}}
\put(1475,603){\rule[-0.175pt]{4.818pt}{0.350pt}}
\put(264,714){\rule[-0.175pt]{4.818pt}{0.350pt}}
\put(242,714){\makebox(0,0)[r]{-0.1}}
\put(1475,714){\rule[-0.175pt]{4.818pt}{0.350pt}}
\put(264,825){\rule[-0.175pt]{4.818pt}{0.350pt}}
\put(242,825){\makebox(0,0)[r]{0}}
\put(1475,825){\rule[-0.175pt]{4.818pt}{0.350pt}}
\put(264,936){\rule[-0.175pt]{4.818pt}{0.350pt}}
\put(242,936){\makebox(0,0)[r]{0.1}}
\put(1475,936){\rule[-0.175pt]{4.818pt}{0.350pt}}
\put(264,158){\rule[-0.175pt]{0.350pt}{4.818pt}}
\put(264,113){\makebox(0,0){0.01}}
\put(264,916){\rule[-0.175pt]{0.350pt}{4.818pt}}
\put(449,158){\rule[-0.175pt]{0.350pt}{2.409pt}}
\put(449,926){\rule[-0.175pt]{0.350pt}{2.409pt}}
\put(558,158){\rule[-0.175pt]{0.350pt}{2.409pt}}
\put(558,926){\rule[-0.175pt]{0.350pt}{2.409pt}}
\put(635,158){\rule[-0.175pt]{0.350pt}{2.409pt}}
\put(635,926){\rule[-0.175pt]{0.350pt}{2.409pt}}
\put(694,158){\rule[-0.175pt]{0.350pt}{2.409pt}}
\put(694,926){\rule[-0.175pt]{0.350pt}{2.409pt}}
\put(743,158){\rule[-0.175pt]{0.350pt}{2.409pt}}
\put(743,926){\rule[-0.175pt]{0.350pt}{2.409pt}}
\put(784,158){\rule[-0.175pt]{0.350pt}{2.409pt}}
\put(784,926){\rule[-0.175pt]{0.350pt}{2.409pt}}
\put(820,158){\rule[-0.175pt]{0.350pt}{2.409pt}}
\put(820,926){\rule[-0.175pt]{0.350pt}{2.409pt}}
\put(851,158){\rule[-0.175pt]{0.350pt}{2.409pt}}
\put(851,926){\rule[-0.175pt]{0.350pt}{2.409pt}}
\put(880,158){\rule[-0.175pt]{0.350pt}{4.818pt}}
\put(880,113){\makebox(0,0){0.1}}
\put(880,916){\rule[-0.175pt]{0.350pt}{4.818pt}}
\put(1065,158){\rule[-0.175pt]{0.350pt}{2.409pt}}
\put(1065,926){\rule[-0.175pt]{0.350pt}{2.409pt}}
\put(1173,158){\rule[-0.175pt]{0.350pt}{2.409pt}}
\put(1173,926){\rule[-0.175pt]{0.350pt}{2.409pt}}
\put(1250,158){\rule[-0.175pt]{0.350pt}{2.409pt}}
\put(1250,926){\rule[-0.175pt]{0.350pt}{2.409pt}}
\put(1310,158){\rule[-0.175pt]{0.350pt}{2.409pt}}
\put(1310,926){\rule[-0.175pt]{0.350pt}{2.409pt}}
\put(1358,158){\rule[-0.175pt]{0.350pt}{2.409pt}}
\put(1358,926){\rule[-0.175pt]{0.350pt}{2.409pt}}
\put(1400,158){\rule[-0.175pt]{0.350pt}{2.409pt}}
\put(1400,926){\rule[-0.175pt]{0.350pt}{2.409pt}}
\put(1435,158){\rule[-0.175pt]{0.350pt}{2.409pt}}
\put(1435,926){\rule[-0.175pt]{0.350pt}{2.409pt}}
\put(1467,158){\rule[-0.175pt]{0.350pt}{2.409pt}}
\put(1467,926){\rule[-0.175pt]{0.350pt}{2.409pt}}
\put(1495,158){\rule[-0.175pt]{0.350pt}{4.818pt}}
\put(1495,113){\makebox(0,0){1}}
\put(1495,916){\rule[-0.175pt]{0.350pt}{4.818pt}}
\put(264,158){\rule[-0.175pt]{296.548pt}{0.350pt}}
\put(1495,158){\rule[-0.175pt]{0.350pt}{187.420pt}}
\put(264,936){\rule[-0.175pt]{296.548pt}{0.350pt}}
\put(45,547){\makebox(0,0)[l]{\shortstack{$ A^{W^+}$}}}
\put(879,68){\makebox(0,0){$x$}}
\put(264,158){\rule[-0.175pt]{0.350pt}{187.420pt}}
\put(1365,871){\makebox(0,0)[r]{$A^{W^+}$}}
\put(1409,871){\raisebox{-1.2pt}{\makebox(0,0){$\Diamond$}}}
\put(387,827){\raisebox{-1.2pt}{\makebox(0,0){$\Diamond$}}}
\put(510,790){\raisebox{-1.2pt}{\makebox(0,0){$\Diamond$}}}
\put(633,749){\raisebox{-1.2pt}{\makebox(0,0){$\Diamond$}}}
\put(756,703){\raisebox{-1.2pt}{\makebox(0,0){$\Diamond$}}}
\put(880,651){\raisebox{-1.2pt}{\makebox(0,0){$\Diamond$}}}
\put(1003,592){\raisebox{-1.2pt}{\makebox(0,0){$\Diamond$}}}
\put(1126,535){\raisebox{-1.2pt}{\makebox(0,0){$\Diamond$}}}
\put(1249,492){\raisebox{-1.2pt}{\makebox(0,0){$\Diamond$}}}
\put(1372,465){\raisebox{-1.2pt}{\makebox(0,0){$\Diamond$}}}
\put(1387,871){\rule[-0.175pt]{15.899pt}{0.350pt}}
\put(1387,861){\rule[-0.175pt]{0.350pt}{4.818pt}}
\put(1453,861){\rule[-0.175pt]{0.350pt}{4.818pt}}
\put(387,797){\rule[-0.175pt]{0.350pt}{14.695pt}}
\put(377,797){\rule[-0.175pt]{4.818pt}{0.350pt}}
\put(377,858){\rule[-0.175pt]{4.818pt}{0.350pt}}
\put(510,762){\rule[-0.175pt]{0.350pt}{13.731pt}}
\put(500,762){\rule[-0.175pt]{4.818pt}{0.350pt}}
\put(500,819){\rule[-0.175pt]{4.818pt}{0.350pt}}
\put(633,721){\rule[-0.175pt]{0.350pt}{13.490pt}}
\put(623,721){\rule[-0.175pt]{4.818pt}{0.350pt}}
\put(623,777){\rule[-0.175pt]{4.818pt}{0.350pt}}
\put(756,674){\rule[-0.175pt]{0.350pt}{13.972pt}}
\put(746,674){\rule[-0.175pt]{4.818pt}{0.350pt}}
\put(746,732){\rule[-0.175pt]{4.818pt}{0.350pt}}
\put(880,620){\rule[-0.175pt]{0.350pt}{14.936pt}}
\put(870,620){\rule[-0.175pt]{4.818pt}{0.350pt}}
\put(870,682){\rule[-0.175pt]{4.818pt}{0.350pt}}
\put(1003,557){\rule[-0.175pt]{0.350pt}{16.863pt}}
\put(993,557){\rule[-0.175pt]{4.818pt}{0.350pt}}
\put(993,627){\rule[-0.175pt]{4.818pt}{0.350pt}}
\put(1126,491){\rule[-0.175pt]{0.350pt}{21.199pt}}
\put(1116,491){\rule[-0.175pt]{4.818pt}{0.350pt}}
\put(1116,579){\rule[-0.175pt]{4.818pt}{0.350pt}}
\put(1249,423){\rule[-0.175pt]{0.350pt}{33.244pt}}
\put(1239,423){\rule[-0.175pt]{4.818pt}{0.350pt}}
\put(1239,561){\rule[-0.175pt]{4.818pt}{0.350pt}}
\put(1372,261){\rule[-0.175pt]{0.350pt}{98.287pt}}
\put(1362,261){\rule[-0.175pt]{4.818pt}{0.350pt}}
\put(1362,669){\rule[-0.175pt]{4.818pt}{0.350pt}}
\end{picture}
\caption{\small\sf
Charged current $e^+ p$ spin asymmetry, Eq. (10) of text, 
for $Q^2=1000$ GeV$^2$, at HERA energies with fully
polarized beams. Statistical errors calculated for 
${\cal L}= 500 \, pb^{-1}$ for each polarization and $\Delta Q^2=500$ GeV$^2$
are shown.
}
\end{figure}
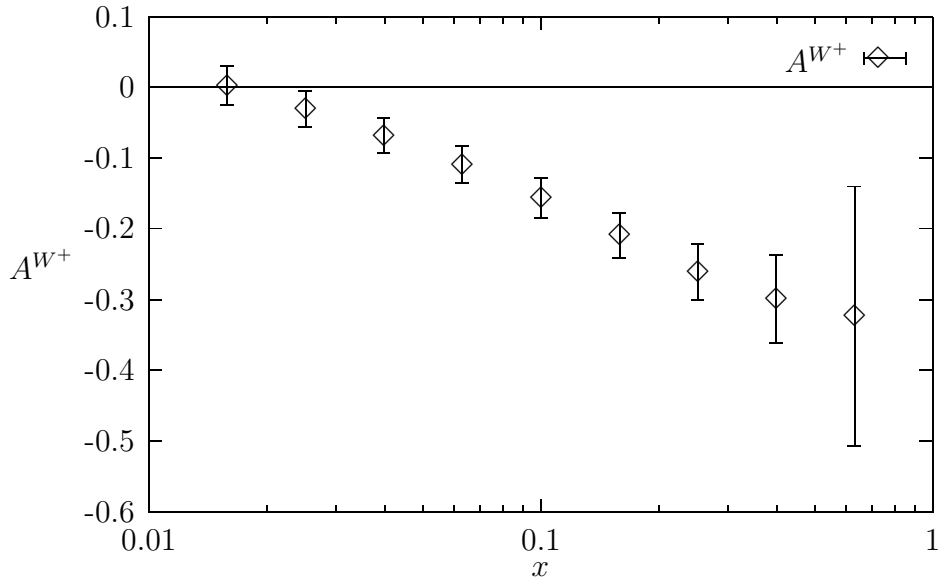
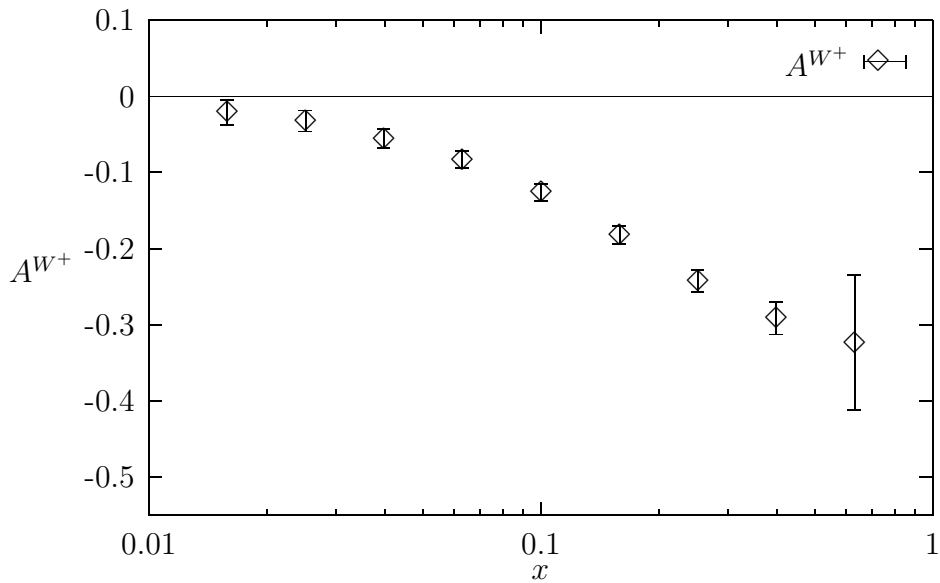
\begin{figure}
% GNUPLOT: LaTeX picture
\setlength{\unitlength}{0.240900pt}
\ifx\plotpoint\undefined\newsavebox{\plotpoint}\fi
\sbox{\plotpoint}{\rule[-0.175pt]{0.350pt}{0.350pt}}%
\begin{picture}(1559,1049)(0,0)
%\tenrm
\sbox{\plotpoint}{\rule[-0.175pt]{0.350pt}{0.350pt}}%
\put(264,816){\rule[-0.175pt]{296.548pt}{0.350pt}}
\put(264,218){\rule[-0.175pt]{4.818pt}{0.350pt}}
\put(242,218){\makebox(0,0)[r]{-0.5}}
\put(1475,218){\rule[-0.175pt]{4.818pt}{0.350pt}}
\put(264,338){\rule[-0.175pt]{4.818pt}{0.350pt}}
\put(242,338){\makebox(0,0)[r]{-0.4}}
\put(1475,338){\rule[-0.175pt]{4.818pt}{0.350pt}}
\put(264,457){\rule[-0.175pt]{4.818pt}{0.350pt}}
\put(242,457){\makebox(0,0)[r]{-0.3}}
\put(1475,457){\rule[-0.175pt]{4.818pt}{0.350pt}}
\put(264,577){\rule[-0.175pt]{4.818pt}{0.350pt}}
\put(242,577){\makebox(0,0)[r]{-0.2}}
\put(1475,577){\rule[-0.175pt]{4.818pt}{0.350pt}}
\put(264,697){\rule[-0.175pt]{4.818pt}{0.350pt}}
\put(242,697){\makebox(0,0)[r]{-0.1}}
\put(1475,697){\rule[-0.175pt]{4.818pt}{0.350pt}}
\put(264,816){\rule[-0.175pt]{4.818pt}{0.350pt}}
\put(242,816){\makebox(0,0)[r]{0}}
\put(1475,816){\rule[-0.175pt]{4.818pt}{0.350pt}}
\put(264,936){\rule[-0.175pt]{4.818pt}{0.350pt}}
\put(242,936){\makebox(0,0)[r]{0.1}}
\put(1475,936){\rule[-0.175pt]{4.818pt}{0.350pt}}
\put(264,158){\rule[-0.175pt]{0.350pt}{4.818pt}}
\put(264,113){\makebox(0,0){0.01}}
\put(264,916){\rule[-0.175pt]{0.350pt}{4.818pt}}
\put(449,158){\rule[-0.175pt]{0.350pt}{2.409pt}}
\put(449,926){\rule[-0.175pt]{0.350pt}{2.409pt}}
\put(558,158){\rule[-0.175pt]{0.350pt}{2.409pt}}
\put(558,926){\rule[-0.175pt]{0.350pt}{2.409pt}}
\put(635,158){\rule[-0.175pt]{0.350pt}{2.409pt}}
\put(635,926){\rule[-0.175pt]{0.350pt}{2.409pt}}
\put(694,158){\rule[-0.175pt]{0.350pt}{2.409pt}}
\put(694,926){\rule[-0.175pt]{0.350pt}{2.409pt}}
\put(743,158){\rule[-0.175pt]{0.350pt}{2.409pt}}
\put(743,926){\rule[-0.175pt]{0.350pt}{2.409pt}}
\put(784,158){\rule[-0.175pt]{0.350pt}{2.409pt}}
\put(784,926){\rule[-0.175pt]{0.350pt}{2.409pt}}
\put(820,158){\rule[-0.175pt]{0.350pt}{2.409pt}}
\put(820,926){\rule[-0.175pt]{0.350pt}{2.409pt}}
\put(851,158){\rule[-0.175pt]{0.350pt}{2.409pt}}
\put(851,926){\rule[-0.175pt]{0.350pt}{2.409pt}}
\put(880,158){\rule[-0.175pt]{0.350pt}{4.818pt}}
\put(880,113){\makebox(0,0){0.1}}
\put(880,916){\rule[-0.175pt]{0.350pt}{4.818pt}}
\put(1065,158){\rule[-0.175pt]{0.350pt}{2.409pt}}
\put(1065,926){\rule[-0.175pt]{0.350pt}{2.409pt}}
\put(1173,158){\rule[-0.175pt]{0.350pt}{2.409pt}}
\put(1173,926){\rule[-0.175pt]{0.350pt}{2.409pt}}
\put(1250,158){\rule[-0.175pt]{0.350pt}{2.409pt}}
\put(1250,926){\rule[-0.175pt]{0.350pt}{2.409pt}}
\put(1310,158){\rule[-0.175pt]{0.350pt}{2.409pt}}
\put(1310,926){\rule[-0.175pt]{0.350pt}{2.409pt}}
\put(1358,158){\rule[-0.175pt]{0.350pt}{2.409pt}}
\put(1358,926){\rule[-0.175pt]{0.350pt}{2.409pt}}
\put(1400,158){\rule[-0.175pt]{0.350pt}{2.409pt}}
\put(1400,926){\rule[-0.175pt]{0.350pt}{2.409pt}}
\put(1435,158){\rule[-0.175pt]{0.350pt}{2.409pt}}
\put(1435,926){\rule[-0.175pt]{0.350pt}{2.409pt}}
\put(1467,158){\rule[-0.175pt]{0.350pt}{2.409pt}}
\put(1467,926){\rule[-0.175pt]{0.350pt}{2.409pt}}
\put(1495,158){\rule[-0.175pt]{0.350pt}{4.818pt}}
\put(1495,113){\makebox(0,0){1}}
\put(1495,916){\rule[-0.175pt]{0.350pt}{4.818pt}}
\put(264,158){\rule[-0.175pt]{296.548pt}{0.350pt}}
\put(1495,158){\rule[-0.175pt]{0.350pt}{187.420pt}}
\put(264,936){\rule[-0.175pt]{296.548pt}{0.350pt}}
\put(45,547){\makebox(0,0)[l]{\shortstack{$ A^{W^+}$}}}
\put(879,68){\makebox(0,0){$x$}}
\put(264,158){\rule[-0.175pt]{0.350pt}{187.420pt}}
\put(1365,871){\makebox(0,0)[r]{$A^{W^+}$}}
\put(1409,871){\raisebox{-1.2pt}{\makebox(0,0){$\Diamond$}}}
\put(387,791){\raisebox{-1.2pt}{\makebox(0,0){$\Diamond$}}}
\put(510,778){\raisebox{-1.2pt}{\makebox(0,0){$\Diamond$}}}
\put(633,750){\raisebox{-1.2pt}{\makebox(0,0){$\Diamond$}}}
\put(756,717){\raisebox{-1.2pt}{\makebox(0,0){$\Diamond$}}}
\put(880,666){\raisebox{-1.2pt}{\makebox(0,0){$\Diamond$}}}
\put(1003,599){\raisebox{-1.2pt}{\makebox(0,0){$\Diamond$}}}
\put(1126,526){\raisebox{-1.2pt}{\makebox(0,0){$\Diamond$}}}
\put(1249,468){\raisebox{-1.2pt}{\makebox(0,0){$\Diamond$}}}
\put(1372,429){\raisebox{-1.2pt}{\makebox(0,0){$\Diamond$}}}
\put(1387,871){\rule[-0.175pt]{15.899pt}{0.350pt}}
\put(1387,861){\rule[-0.175pt]{0.350pt}{4.818pt}}
\put(1453,861){\rule[-0.175pt]{0.350pt}{4.818pt}}
\put(387,771){\rule[-0.175pt]{0.350pt}{9.636pt}}
\put(377,771){\rule[-0.175pt]{4.818pt}{0.350pt}}
\put(377,811){\rule[-0.175pt]{4.818pt}{0.350pt}}
\put(510,761){\rule[-0.175pt]{0.350pt}{7.950pt}}
\put(500,761){\rule[-0.175pt]{4.818pt}{0.350pt}}
\put(500,794){\rule[-0.175pt]{4.818pt}{0.350pt}}
\put(633,735){\rule[-0.175pt]{0.350pt}{7.227pt}}
\put(623,735){\rule[-0.175pt]{4.818pt}{0.350pt}}
\put(623,765){\rule[-0.175pt]{4.818pt}{0.350pt}}
\put(756,703){\rule[-0.175pt]{0.350pt}{6.504pt}}
\put(746,703){\rule[-0.175pt]{4.818pt}{0.350pt}}
\put(746,730){\rule[-0.175pt]{4.818pt}{0.350pt}}
\put(880,652){\rule[-0.175pt]{0.350pt}{6.504pt}}
\put(870,652){\rule[-0.175pt]{4.818pt}{0.350pt}}
\put(870,679){\rule[-0.175pt]{4.818pt}{0.350pt}}
\put(1003,584){\rule[-0.175pt]{0.350pt}{6.986pt}}
\put(993,584){\rule[-0.175pt]{4.818pt}{0.350pt}}
\put(993,613){\rule[-0.175pt]{4.818pt}{0.350pt}}
\put(1126,509){\rule[-0.175pt]{0.350pt}{8.191pt}}
\put(1116,509){\rule[-0.175pt]{4.818pt}{0.350pt}}
\put(1116,543){\rule[-0.175pt]{4.818pt}{0.350pt}}
\put(1249,442){\rule[-0.175pt]{0.350pt}{12.286pt}}
\put(1239,442){\rule[-0.175pt]{4.818pt}{0.350pt}}
\put(1239,493){\rule[-0.175pt]{4.818pt}{0.350pt}}
\put(1372,323){\rule[-0.175pt]{0.350pt}{51.071pt}}
\put(1362,323){\rule[-0.175pt]{4.818pt}{0.350pt}}
\put(1362,535){\rule[-0.175pt]{4.818pt}{0.350pt}}
\end{picture}
\caption{\small\sf Charged current $e^+ p$ spin asymmetry, Eq. (10) of text, 
averaged over the allowed $Q^2$ region, at HERA energies with fully
polarized beams. Statistical errors calculated for 
${\cal L}= 500 \, pb^{-1}$ for each polarization are shown.
}
\end{figure}
\begin{figure}
% GNUPLOT: LaTeX picture
\setlength{\unitlength}{0.240900pt}
\ifx\plotpoint\undefined\newsavebox{\plotpoint}\fi
\sbox{\plotpoint}{\rule[-0.175pt]{0.350pt}{0.350pt}}%
\begin{picture}(1559,1049)(0,0)
\sbox{\plotpoint}{\rule[-0.175pt]{0.350pt}{0.350pt}}%
\put(264,527){\rule[-0.175pt]{296.548pt}{0.350pt}}
\put(264,199){\rule[-0.175pt]{4.818pt}{0.350pt}}
\put(242,199){\makebox(0,0)[r]{-0.04}}
\put(1475,199){\rule[-0.175pt]{4.818pt}{0.350pt}}
\put(264,281){\rule[-0.175pt]{4.818pt}{0.350pt}}
\put(242,281){\makebox(0,0)[r]{-0.03}}
\put(1475,281){\rule[-0.175pt]{4.818pt}{0.350pt}}
\put(264,363){\rule[-0.175pt]{4.818pt}{0.350pt}}
\put(242,363){\makebox(0,0)[r]{-0.02}}
\put(1475,363){\rule[-0.175pt]{4.818pt}{0.350pt}}
\put(264,445){\rule[-0.175pt]{4.818pt}{0.350pt}}
\put(242,445){\makebox(0,0)[r]{-0.01}}
\put(1475,445){\rule[-0.175pt]{4.818pt}{0.350pt}}
\put(264,527){\rule[-0.175pt]{4.818pt}{0.350pt}}
\put(242,527){\makebox(0,0)[r]{0}}
\put(1475,527){\rule[-0.175pt]{4.818pt}{0.350pt}}
\put(264,608){\rule[-0.175pt]{4.818pt}{0.350pt}}
\put(242,608){\makebox(0,0)[r]{0.01}}
\put(1475,608){\rule[-0.175pt]{4.818pt}{0.350pt}}
\put(264,690){\rule[-0.175pt]{4.818pt}{0.350pt}}
\put(242,690){\makebox(0,0)[r]{0.02}}
\put(1475,690){\rule[-0.175pt]{4.818pt}{0.350pt}}
\put(264,772){\rule[-0.175pt]{4.818pt}{0.350pt}}
\put(242,772){\makebox(0,0)[r]{0.03}}
\put(1475,772){\rule[-0.175pt]{4.818pt}{0.350pt}}
\put(264,854){\rule[-0.175pt]{4.818pt}{0.350pt}}
\put(242,854){\makebox(0,0)[r]{0.04}}
\put(1475,854){\rule[-0.175pt]{4.818pt}{0.350pt}}
\put(264,936){\rule[-0.175pt]{4.818pt}{0.350pt}}
\put(242,936){\makebox(0,0)[r]{0.05}}
\put(1475,936){\rule[-0.175pt]{4.818pt}{0.350pt}}
\put(264,158){\rule[-0.175pt]{0.350pt}{2.409pt}}
\put(264,926){\rule[-0.175pt]{0.350pt}{2.409pt}}
\put(294,158){\rule[-0.175pt]{0.350pt}{2.409pt}}
\put(294,926){\rule[-0.175pt]{0.350pt}{2.409pt}}
\put(321,158){\rule[-0.175pt]{0.350pt}{4.818pt}}
\put(321,113){\makebox(0,0){0.01}}
\put(321,916){\rule[-0.175pt]{0.350pt}{4.818pt}}
\put(498,158){\rule[-0.175pt]{0.350pt}{2.409pt}}
\put(498,926){\rule[-0.175pt]{0.350pt}{2.409pt}}
\put(601,158){\rule[-0.175pt]{0.350pt}{2.409pt}}
\put(601,926){\rule[-0.175pt]{0.350pt}{2.409pt}}
\put(674,158){\rule[-0.175pt]{0.350pt}{2.409pt}}
\put(674,926){\rule[-0.175pt]{0.350pt}{2.409pt}}
\put(731,158){\rule[-0.175pt]{0.350pt}{2.409pt}}
\put(731,926){\rule[-0.175pt]{0.350pt}{2.409pt}}
\put(778,158){\rule[-0.175pt]{0.350pt}{2.409pt}}
\put(778,926){\rule[-0.175pt]{0.350pt}{2.409pt}}
\put(817,158){\rule[-0.175pt]{0.350pt}{2.409pt}}
\put(817,926){\rule[-0.175pt]{0.350pt}{2.409pt}}
\put(851,158){\rule[-0.175pt]{0.350pt}{2.409pt}}
\put(851,926){\rule[-0.175pt]{0.350pt}{2.409pt}}
\put(881,158){\rule[-0.175pt]{0.350pt}{2.409pt}}
\put(881,926){\rule[-0.175pt]{0.350pt}{2.409pt}}
\put(908,158){\rule[-0.175pt]{0.350pt}{4.818pt}}
\put(908,113){\makebox(0,0){0.1}}
\put(908,916){\rule[-0.175pt]{0.350pt}{4.818pt}}
\put(1085,158){\rule[-0.175pt]{0.350pt}{2.409pt}}
\put(1085,926){\rule[-0.175pt]{0.350pt}{2.409pt}}
\put(1188,158){\rule[-0.175pt]{0.350pt}{2.409pt}}
\put(1188,926){\rule[-0.175pt]{0.350pt}{2.409pt}}
\put(1261,158){\rule[-0.175pt]{0.350pt}{2.409pt}}
\put(1261,926){\rule[-0.175pt]{0.350pt}{2.409pt}}
\put(1318,158){\rule[-0.175pt]{0.350pt}{2.409pt}}
\put(1318,926){\rule[-0.175pt]{0.350pt}{2.409pt}}
\put(1365,158){\rule[-0.175pt]{0.350pt}{2.409pt}}
\put(1365,926){\rule[-0.175pt]{0.350pt}{2.409pt}}
\put(1404,158){\rule[-0.175pt]{0.350pt}{2.409pt}}
\put(1404,926){\rule[-0.175pt]{0.350pt}{2.409pt}}
\put(1438,158){\rule[-0.175pt]{0.350pt}{2.409pt}}
\put(1438,926){\rule[-0.175pt]{0.350pt}{2.409pt}}
\put(1468,158){\rule[-0.175pt]{0.350pt}{2.409pt}}
\put(1468,926){\rule[-0.175pt]{0.350pt}{2.409pt}}
\put(1495,158){\rule[-0.175pt]{0.350pt}{4.818pt}}
\put(1495,113){\makebox(0,0){1}}
\put(1495,916){\rule[-0.175pt]{0.350pt}{4.818pt}}
\put(264,158){\rule[-0.175pt]{296.548pt}{0.350pt}}
\put(1495,158){\rule[-0.175pt]{0.350pt}{187.420pt}}
\put(264,936){\rule[-0.175pt]{296.548pt}{0.350pt}}
\put(45,547){\makebox(0,0)[l]{\shortstack{$ A^{e^- p}_{nc}$}}}
\put(879,68){\makebox(0,0){$x$}}
\put(264,158){\rule[-0.175pt]{0.350pt}{187.420pt}}
\put(1365,871){\makebox(0,0)[r]{$A^{e^- p}_{nc}$}}
\put(1409,871){\raisebox{-1.2pt}{\makebox(0,0){$\Diamond$}}}
\put(438,514){\raisebox{-1.2pt}{\makebox(0,0){$\Diamond$}}}
\put(556,511){\raisebox{-1.2pt}{\makebox(0,0){$\Diamond$}}}
\put(673,509){\raisebox{-1.2pt}{\makebox(0,0){$\Diamond$}}}
\put(791,506){\raisebox{-1.2pt}{\makebox(0,0){$\Diamond$}}}
\put(908,503){\raisebox{-1.2pt}{\makebox(0,0){$\Diamond$}}}
\put(1025,499){\raisebox{-1.2pt}{\makebox(0,0){$\Diamond$}}}
\put(1143,495){\raisebox{-1.2pt}{\makebox(0,0){$\Diamond$}}}
\put(1260,493){\raisebox{-1.2pt}{\makebox(0,0){$\Diamond$}}}
\put(1378,491){\raisebox{-1.2pt}{\makebox(0,0){$\Diamond$}}}
\put(1387,871){\rule[-0.175pt]{15.899pt}{0.350pt}}
\put(1387,861){\rule[-0.175pt]{0.350pt}{4.818pt}}
\put(1453,861){\rule[-0.175pt]{0.350pt}{4.818pt}}
\put(438,456){\rule[-0.175pt]{0.350pt}{27.944pt}}
\put(428,456){\rule[-0.175pt]{4.818pt}{0.350pt}}
\put(428,572){\rule[-0.175pt]{4.818pt}{0.350pt}}
\put(556,455){\rule[-0.175pt]{0.350pt}{26.981pt}}
\put(546,455){\rule[-0.175pt]{4.818pt}{0.350pt}}
\put(546,567){\rule[-0.175pt]{4.818pt}{0.350pt}}
\put(673,453){\rule[-0.175pt]{0.350pt}{26.740pt}}
\put(663,453){\rule[-0.175pt]{4.818pt}{0.350pt}}
\put(663,564){\rule[-0.175pt]{4.818pt}{0.350pt}}
\put(791,450){\rule[-0.175pt]{0.350pt}{26.981pt}}
\put(781,450){\rule[-0.175pt]{4.818pt}{0.350pt}}
\put(781,562){\rule[-0.175pt]{4.818pt}{0.350pt}}
\put(908,445){\rule[-0.175pt]{0.350pt}{27.703pt}}
\put(898,445){\rule[-0.175pt]{4.818pt}{0.350pt}}
\put(898,560){\rule[-0.175pt]{4.818pt}{0.350pt}}
\put(1025,438){\rule[-0.175pt]{0.350pt}{29.149pt}}
\put(1015,438){\rule[-0.175pt]{4.818pt}{0.350pt}}
\put(1015,559){\rule[-0.175pt]{4.818pt}{0.350pt}}
\put(1143,427){\rule[-0.175pt]{0.350pt}{33.003pt}}
\put(1133,427){\rule[-0.175pt]{4.818pt}{0.350pt}}
\put(1133,564){\rule[-0.175pt]{4.818pt}{0.350pt}}
\put(1260,402){\rule[-0.175pt]{0.350pt}{43.844pt}}
\put(1250,402){\rule[-0.175pt]{4.818pt}{0.350pt}}
\put(1250,584){\rule[-0.175pt]{4.818pt}{0.350pt}}
\put(1378,197){\rule[-0.175pt]{0.350pt}{141.649pt}}
\put(1368,197){\rule[-0.175pt]{4.818pt}{0.350pt}}
\put(1368,785){\rule[-0.175pt]{4.818pt}{0.350pt}}
\end{picture}
\caption{\small\sf Neutral  current $e^- p$ single spin 
asymmetry, Eq.(11) of text,  for
$Q^2=1000$ GeV$^2$  at HERA energies with fully polarized proton beams.
The statistical errors calculated for ${\cal L}= 500 \, pb^{-1}$ 
for each polarization and $\Delta Q^2=500$ GeV$^2$ are shown.
}
\end{figure}
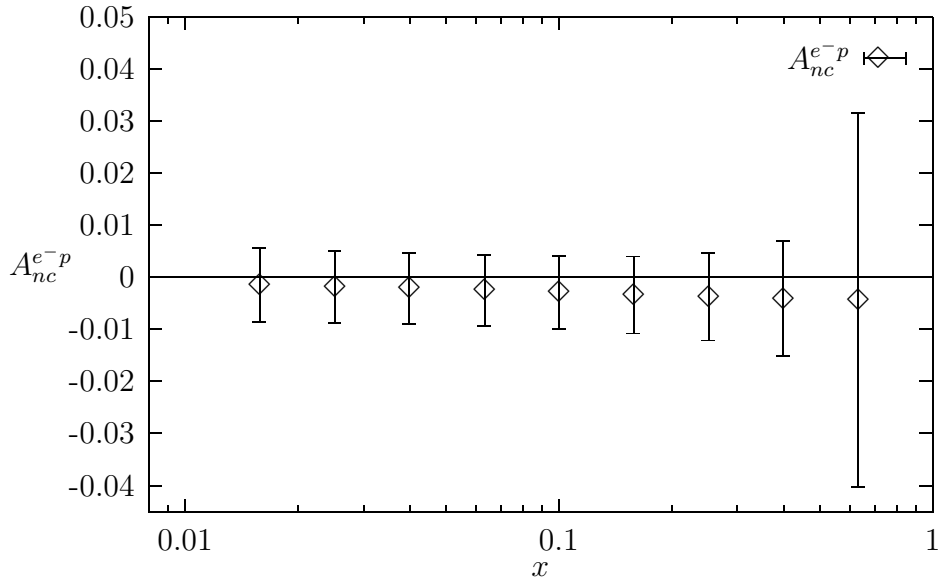
\begin{figure}
% GNUPLOT: LaTeX picture
\setlength{\unitlength}{0.240900pt}
\ifx\plotpoint\undefined\newsavebox{\plotpoint}\fi
\sbox{\plotpoint}{\rule[-0.175pt]{0.350pt}{0.350pt}}%
\begin{picture}(1559,1049)(0,0)
%\tenrm
\sbox{\plotpoint}{\rule[-0.175pt]{0.350pt}{0.350pt}}%
\put(264,577){\rule[-0.175pt]{296.548pt}{0.350pt}}
\put(264,218){\rule[-0.175pt]{4.818pt}{0.350pt}}
\put(242,218){\makebox(0,0)[r]{-0.03}}
\put(1475,218){\rule[-0.175pt]{4.818pt}{0.350pt}}
\put(264,338){\rule[-0.175pt]{4.818pt}{0.350pt}}
\put(242,338){\makebox(0,0)[r]{-0.02}}
\put(1475,338){\rule[-0.175pt]{4.818pt}{0.350pt}}
\put(264,457){\rule[-0.175pt]{4.818pt}{0.350pt}}
\put(242,457){\makebox(0,0)[r]{-0.01}}
\put(1475,457){\rule[-0.175pt]{4.818pt}{0.350pt}}
\put(264,577){\rule[-0.175pt]{4.818pt}{0.350pt}}
\put(242,577){\makebox(0,0)[r]{0}}
\put(1475,577){\rule[-0.175pt]{4.818pt}{0.350pt}}
\put(264,697){\rule[-0.175pt]{4.818pt}{0.350pt}}
\put(242,697){\makebox(0,0)[r]{0.01}}
\put(1475,697){\rule[-0.175pt]{4.818pt}{0.350pt}}
\put(264,816){\rule[-0.175pt]{4.818pt}{0.350pt}}
\put(242,816){\makebox(0,0)[r]{0.02}}
\put(1475,816){\rule[-0.175pt]{4.818pt}{0.350pt}}
\put(264,936){\rule[-0.175pt]{4.818pt}{0.350pt}}
\put(242,936){\makebox(0,0)[r]{0.03}}
\put(1475,936){\rule[-0.175pt]{4.818pt}{0.350pt}}
\put(264,158){\rule[-0.175pt]{0.350pt}{2.409pt}}
\put(264,926){\rule[-0.175pt]{0.350pt}{2.409pt}}
\put(294,158){\rule[-0.175pt]{0.350pt}{2.409pt}}
\put(294,926){\rule[-0.175pt]{0.350pt}{2.409pt}}
\put(321,158){\rule[-0.175pt]{0.350pt}{4.818pt}}
\put(321,113){\makebox(0,0){0.01}}
\put(321,916){\rule[-0.175pt]{0.350pt}{4.818pt}}
\put(498,158){\rule[-0.175pt]{0.350pt}{2.409pt}}
\put(498,926){\rule[-0.175pt]{0.350pt}{2.409pt}}
\put(601,158){\rule[-0.175pt]{0.350pt}{2.409pt}}
\put(601,926){\rule[-0.175pt]{0.350pt}{2.409pt}}
\put(674,158){\rule[-0.175pt]{0.350pt}{2.409pt}}
\put(674,926){\rule[-0.175pt]{0.350pt}{2.409pt}}
\put(731,158){\rule[-0.175pt]{0.350pt}{2.409pt}}
\put(731,926){\rule[-0.175pt]{0.350pt}{2.409pt}}
\put(778,158){\rule[-0.175pt]{0.350pt}{2.409pt}}
\put(778,926){\rule[-0.175pt]{0.350pt}{2.409pt}}
\put(817,158){\rule[-0.175pt]{0.350pt}{2.409pt}}
\put(817,926){\rule[-0.175pt]{0.350pt}{2.409pt}}
\put(851,158){\rule[-0.175pt]{0.350pt}{2.409pt}}
\put(851,926){\rule[-0.175pt]{0.350pt}{2.409pt}}
\put(881,158){\rule[-0.175pt]{0.350pt}{2.409pt}}
\put(881,926){\rule[-0.175pt]{0.350pt}{2.409pt}}
\put(908,158){\rule[-0.175pt]{0.350pt}{4.818pt}}
\put(908,113){\makebox(0,0){0.1}}
\put(908,916){\rule[-0.175pt]{0.350pt}{4.818pt}}
\put(1085,158){\rule[-0.175pt]{0.350pt}{2.409pt}}
\put(1085,926){\rule[-0.175pt]{0.350pt}{2.409pt}}
\put(1188,158){\rule[-0.175pt]{0.350pt}{2.409pt}}
\put(1188,926){\rule[-0.175pt]{0.350pt}{2.409pt}}
\put(1261,158){\rule[-0.175pt]{0.350pt}{2.409pt}}
\put(1261,926){\rule[-0.175pt]{0.350pt}{2.409pt}}
\put(1318,158){\rule[-0.175pt]{0.350pt}{2.409pt}}
\put(1318,926){\rule[-0.175pt]{0.350pt}{2.409pt}}
\put(1365,158){\rule[-0.175pt]{0.350pt}{2.409pt}}
\put(1365,926){\rule[-0.175pt]{0.350pt}{2.409pt}}
\put(1404,158){\rule[-0.175pt]{0.350pt}{2.409pt}}
\put(1404,926){\rule[-0.175pt]{0.350pt}{2.409pt}}
\put(1438,158){\rule[-0.175pt]{0.350pt}{2.409pt}}
\put(1438,926){\rule[-0.175pt]{0.350pt}{2.409pt}}
\put(1468,158){\rule[-0.175pt]{0.350pt}{2.409pt}}
\put(1468,926){\rule[-0.175pt]{0.350pt}{2.409pt}}
\put(1495,158){\rule[-0.175pt]{0.350pt}{4.818pt}}
\put(1495,113){\makebox(0,0){1}}
\put(1495,916){\rule[-0.175pt]{0.350pt}{4.818pt}}
\put(264,158){\rule[-0.175pt]{296.548pt}{0.350pt}}
\put(1495,158){\rule[-0.175pt]{0.350pt}{187.420pt}}
\put(264,936){\rule[-0.175pt]{296.548pt}{0.350pt}}
\put(45,547){\makebox(0,0)[l]{\shortstack{$ A^{e^- p}_{nc}$}}}
\put(879,68){\makebox(0,0){$x$}}
\put(264,158){\rule[-0.175pt]{0.350pt}{187.420pt}}
\put(1365,871){\makebox(0,0)[r]{$A^{e^- p}_{nc}$}}
\put(1409,871){\raisebox{-1.2pt}{\makebox(0,0){$\Diamond$}}}
\put(321,570){\raisebox{-1.2pt}{\makebox(0,0){$\Diamond$}}}
\put(438,563){\raisebox{-1.2pt}{\makebox(0,0){$\Diamond$}}}
\put(556,555){\raisebox{-1.2pt}{\makebox(0,0){$\Diamond$}}}
\put(673,545){\raisebox{-1.2pt}{\makebox(0,0){$\Diamond$}}}
\put(791,532){\raisebox{-1.2pt}{\makebox(0,0){$\Diamond$}}}
\put(908,519){\raisebox{-1.2pt}{\makebox(0,0){$\Diamond$}}}
\put(1025,507){\raisebox{-1.2pt}{\makebox(0,0){$\Diamond$}}}
\put(1143,499){\raisebox{-1.2pt}{\makebox(0,0){$\Diamond$}}}
\put(1260,500){\raisebox{-1.2pt}{\makebox(0,0){$\Diamond$}}}
\put(1378,509){\raisebox{-1.2pt}{\makebox(0,0){$\Diamond$}}}
\put(1387,871){\rule[-0.175pt]{15.899pt}{0.350pt}}
\put(1387,861){\rule[-0.175pt]{0.350pt}{4.818pt}}
\put(1453,861){\rule[-0.175pt]{0.350pt}{4.818pt}}
\put(321,496){\rule[-0.175pt]{0.350pt}{35.412pt}}
\put(311,496){\rule[-0.175pt]{4.818pt}{0.350pt}}
\put(311,643){\rule[-0.175pt]{4.818pt}{0.350pt}}
\put(438,507){\rule[-0.175pt]{0.350pt}{26.981pt}}
\put(428,507){\rule[-0.175pt]{4.818pt}{0.350pt}}
\put(428,619){\rule[-0.175pt]{4.818pt}{0.350pt}}
\put(556,505){\rule[-0.175pt]{0.350pt}{23.849pt}}
\put(546,505){\rule[-0.175pt]{4.818pt}{0.350pt}}
\put(546,604){\rule[-0.175pt]{4.818pt}{0.350pt}}
\put(673,498){\rule[-0.175pt]{0.350pt}{22.885pt}}
\put(663,498){\rule[-0.175pt]{4.818pt}{0.350pt}}
\put(663,593){\rule[-0.175pt]{4.818pt}{0.350pt}}
\put(791,486){\rule[-0.175pt]{0.350pt}{22.163pt}}
\put(781,486){\rule[-0.175pt]{4.818pt}{0.350pt}}
\put(781,578){\rule[-0.175pt]{4.818pt}{0.350pt}}
\put(908,472){\rule[-0.175pt]{0.350pt}{22.404pt}}
\put(898,472){\rule[-0.175pt]{4.818pt}{0.350pt}}
\put(898,565){\rule[-0.175pt]{4.818pt}{0.350pt}}
\put(1025,458){\rule[-0.175pt]{0.350pt}{23.608pt}}
\put(1015,458){\rule[-0.175pt]{4.818pt}{0.350pt}}
\put(1015,556){\rule[-0.175pt]{4.818pt}{0.350pt}}
\put(1143,445){\rule[-0.175pt]{0.350pt}{26.258pt}}
\put(1133,445){\rule[-0.175pt]{4.818pt}{0.350pt}}
\put(1133,554){\rule[-0.175pt]{4.818pt}{0.350pt}}
\put(1260,428){\rule[-0.175pt]{0.350pt}{34.930pt}}
\put(1250,428){\rule[-0.175pt]{4.818pt}{0.350pt}}
\put(1250,573){\rule[-0.175pt]{4.818pt}{0.350pt}}
\put(1378,276){\rule[-0.175pt]{0.350pt}{112.500pt}}
\put(1368,276){\rule[-0.175pt]{4.818pt}{0.350pt}}
\put(1368,743){\rule[-0.175pt]{4.818pt}{0.350pt}}
\end{picture}
\caption{\small\sf Neutral  current $e^- p$ single spin 
asymmetry, Eq.(11) of text,  averaged over the allowed
$Q^2$ region at HERA energies with fully polarized proton beams.
The statistical errors calculated for ${\cal L}= 500 \, pb^{-1}$ 
for each polarization are shown.
}
\end{figure}
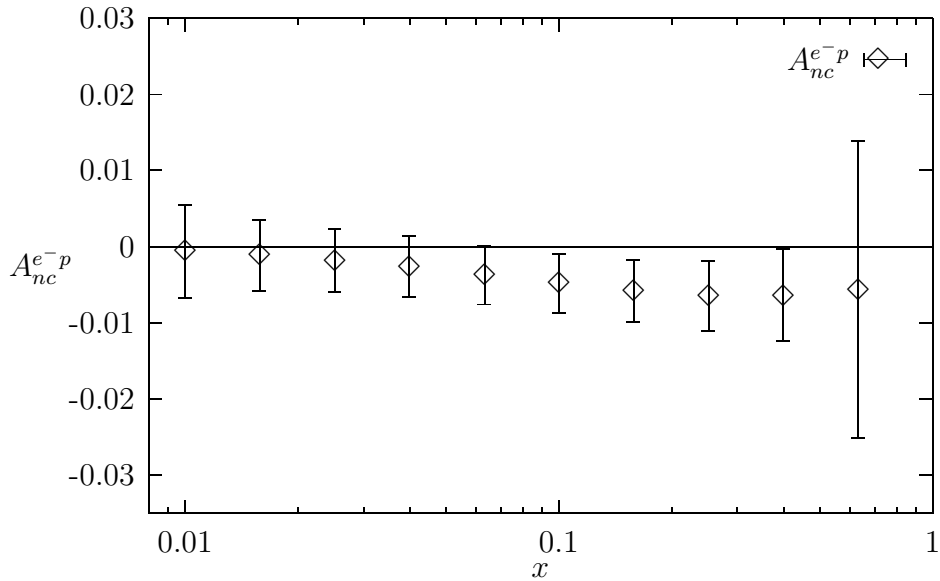
\end{document}